\documentclass{article}

\usepackage{PRIMEarxiv}
\usepackage{amsmath}
\usepackage[utf8]{inputenc} 
\usepackage[T1]{fontenc}    
\usepackage{hyperref}       
\usepackage{url}            
\usepackage{booktabs}       
\usepackage{amsfonts}       
\usepackage{nicefrac}       
\usepackage{microtype}      
\usepackage{lipsum}
\usepackage{fancyhdr}       
\usepackage{graphicx}       
\usepackage{multirow}
\graphicspath{{media/}}     

\title{Which Types of Heterogeneity Matter for Root Cause Localization in Microservice Systems?
}

\author{
  Runzhou Wang \\
  Nankai University \\
  Tianjin,China\\
  \texttt{1120240403@mail.nankai.edu.cn} \\
   \And
  Shenglin Zhang \\
  Nankai University \\
  Tianjin,China\\
  \texttt{zhangsl@nankai.edu.cn} \\
  \AND
  Wenwei Gu \\
  Nankai University \\
  Tianjin,China\\
  \texttt{wwgu@nankai.edu.cn} \\
  \And
  Yongxin Zhao \\
  Nankai University \\
  Tianjin,China\\
  \texttt{zyx\_nkcs@mail.nankai.edu.cn} \\
  \And
  Chenyu Zhao \\
  Nankai University \\
  Tianjin,China\\
  \texttt{zhaochenyu@mail.nankai.edu.cn} \\
  \And
  Dan Pei \\
  Tsinghua University \&BNRist \\
  Beijing,China\\
  \texttt{peidan@tsinghua.edu.cn} \\
  \And
  Yuxuan Chen \\
  Nankai University \\
  Tianjin,China\\
  \texttt{2313193@mail.nankai.edu.cn} \\
  \And
  Yangyuxin Huang \\
  Nankai University \\
  Tianjin,China\\
  \texttt{2213789@mail.nankai.edu.cn} \\
}

\begin{document}
\maketitle

\begin{abstract}

Microservice root cause localization is fundamentally challenged by the inherent heterogeneity of cloud-native systems, which encompasses diverse observability data and multiple system entities. Existing approaches typically focus on only one aspect of heterogeneity and thus fail to capture its full diagnostic value. In this work, we systematically examine the multifaceted role of heterogeneity within both microservice systems and the RCL process. This analysis motivates a deeper investigation into how entity-level distinctions and their asymmetric dependencies influence fault behavior. Our empirical analysis of two microservice benchmarks reveals that entity-level heterogeneity naturally gives rise to heterogeneous fault propagation, which is highly asymmetric and dominated by cross-layer interactions between services and hosts. In light of this, we propose NexusRCL, a semi-supervised framework that internalizes these propagation patterns by formalizing services and hosts as distinct node types within a heterogeneous graph. This design, coupled with an event-based abstraction mechanism, allows NexusRCL to effectively capture both data level and entity-level heterogeneity while minimizing labeling costs through active learning. Comprehensive evaluations on two industrial benchmark datasets demonstrate NexusRCL's superior performance, achieving improvements of up to 49.85\% in Top-1 accuracy (A@1) and 32.70\% in Average Top-5 accuracy (A@5) compared to state-of-the-art methods.
\end{abstract}

\keywords{Microservice Systems\and Heterogeneous Graph\and Root Cause Localization\and Active Learning}
\section{Introduction}

\begin{figure}[t]
    \centering
    \includegraphics[width=0.46\textwidth, trim=0cm 13cm 16cm 0cm, clip]{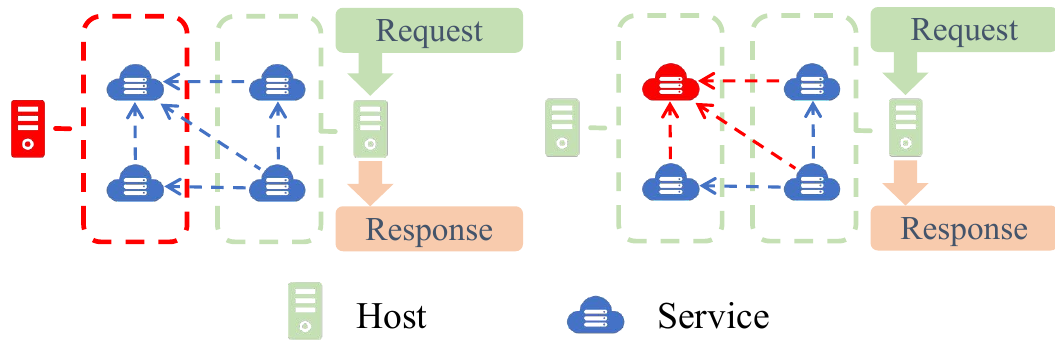}
    \caption{Faults in microservice architectures: host-level fault (left) and service-level fault (right).}
    \label{fig:fault_cases}
\end{figure}

Microservice architectures decompose complex applications into small, loosely coupled services, enhancing flexibility, scalability, and deployment efficiency in cloud-native systems~\cite{kratzke2017understanding}. 
However, the inherent complexity and dynamism of microservice systems make faults unavoidable. Faults may arise at the host or service layer and often lead to service interruptions (see Fig.~\ref{fig:fault_cases})~\cite{soylemez2022challenges}. 
Moreover, service dependencies and deployment relationships between services and hosts often facilitate cascading failures, further compromising system reliability and underscoring the need for effective fault management in microservice systems~\cite{souza2024dependable,jayalath2024microservice}.

These challenges in fault management require the rapid identification of fault origins to enable effective recovery and prevent further propagation. Root Cause Localization (RCL) addresses this need by accurately pinpointing fault sources, facilitating targeted mitigation~\cite{liu2021microhecl, wu2020microrca, ikram2022root}. Traditional RCL methods leverage metrics, logs, and traces, employing techniques such as Bayesian networks~\cite{zhang2021cloudrca} and random walks~\cite{wu2020microrca, wang2018cloudranger} to infer potential root causes~\cite{wittkopp2024logrca, zhang2024trace, gu2023trinityrcl}.

Recently, Graph Neural Networks (GNNs) have emerged as an effective approach for RCL by explicitly modeling service dependencies and fault propagation \textbf{between services}~\cite{lee2023eadro,yu2023nezha,zheng2024mulan}.
Motivated by the inherent heterogeneity of microservice systems, recent RCL studies further adopt heterogeneous graph modeling to explicitly differentiate data sources and service interaction~\cite{tao2024diagnosing,yang2025hg,zhao2024chase}.
For example, MicroDig~\cite{tao2024diagnosing} uses heterogeneous propagation graphs with random walks, while ART~\cite{sun2024art} and DeepHunt~\cite{sun2025interpretable} employ heterogeneous graph structures for unsupervised RCL. HG-PAD~\cite{yang2025hg} models dynamic dependencies among heterogeneous services using graph attention mechanisms.

Despite these advances, existing heterogeneous graph-based RCL methods still face fundamental limitations when deployed in real-world microservice systems. A defining characteristic of such systems lies in \textbf{entity-level heterogeneity}: different components, typically physical infrastructure (hosts) and virtual components (services), vary not only in observable features, but also in semantics, operational roles, and fault behaviors. However, existing approaches either ignore this heterogeneity or handle heterogeneous entities in an overly uniform manner, leading to two critical drawbacks that limit their real-world applicability.

\textbf{First, obscured entity distinctions hinder accurate attribution.}
Many existing RCL methods emphasize monitoring data heterogeneity and directly fuse multi-source monitoring signals into unified representations, without explicitly modeling system entities. Consequently, service- and host-level signals become entangled, obscuring semantic boundaries and weakening entity-specific fault attribution.

\textbf{Second, asymmetric cross-layer fault propagation is not sufficiently modeled.}  
Most approaches assume uniform or symmetric interactions across entities, whereas fault propagation in microservice systems is inherently asymmetric: service-level faults often propagate across layers, while reverse propagation is rare. Ignoring this structural asymmetry limits the effectiveness of fault diffusion modeling in practice.

These observations motivate us to view heterogeneous graphs not merely as a modeling convenience, but as a necessary representation for capturing the diverse forms of heterogeneity inherent in microservice systems. Through empirical analysis of two microservice benchmarks, we uncover dominant fault propagation patterns and identify two main sources of heterogeneity. This naturally gives rise to our core research question: \emph{Which types of heterogeneity matter for RCL in microservice systems?}

Building on the empirical analysis and concrete examples presented in Section~\ref{sec:MOTIVATION}, we argue that both \textbf{data-level} heterogeneity, originating from diverse observability modalities, and \textbf{entity-level} heterogeneity, reflecting the distinct roles of hosts and services, are essential for effective RCL. Despite the promise of heterogeneous graph modeling, fully exploiting such graphs for RCL in microservice systems remains challenging.

\textbf{Challenge 1: Modeling inter-layer heterogeneity under dynamic deployment.}  
Service-to-host relationships continuously change due to auto-scaling and container migration, complicating inter-layer dependency modeling. Accurate capture of such dynamics, together with heterogeneous observability signals, is essential for characterizing fault propagation.

\textbf{Challenge 2: Balance between labeling cost and localization performance.}  
Supervised RCL methods achieve strong performance but require labeled fault cases that are expensive to collect at scale. Unsupervised approaches reduce annotation costs, yet their reliance on assumptions such as metric stationarity often breaks down in real-world systems, making accurate localization under limited labels an open problem.

To address these challenges, we propose \textbf{NexusRCL}, a semi-supervised RCL framework that integrates \textbf{explicit heterogeneous graph modeling} with \textbf{active learning}. As illustrated in Fig.~\ref{fig:NexusRCL_overview}, NexusRCL consists of two key components.

\textbf{Layer-Aware Heterogeneous Graph Modeling.}  
To address Challenge~1, NexusRCL models hosts and services as distinct node types and captures their interactions through typed edges. Unlike prior work that only considers data-source heterogeneity or treats hosts and services as homogeneous nodes, NexusRCL formalizes entity-level heterogeneity by separating hosts and services in the graph. To reduce instability from auto-scaling and container churn, NexusRCL abstracts runtime events at the service level instead of modeling volatile components such as pods. This design preserves semantic consistency and enables robust cross-layer fault propagation modeling.

\textbf{Semi-Supervised Active Learning.}  
To address Challenge~2, NexusRCL employs a three-step active learning strategy. It first clusters heterogeneous graph embeddings, then labels cluster medoids, propagates pseudo-labels within clusters, and finally queries samples with high predictive uncertainty. This strategy substantially reduces labeling cost while maintaining strong localization performance.

In summary, NexusRCL provides a practical solution for RCL in microservice systems by jointly modeling entity-level and data-level heterogeneity while improving label efficiency. The main contributions of this work are summarized as follows:

\begin{enumerate}
    \item We identify fundamental limitations of existing heterogeneous graph-based RCL methods and, through an empirical study, highlight the importance of explicitly modeling entity-level heterogeneity in microservice systems.
    \item We propose a semi-supervised active learning strategy that substantially reduces labeling cost while preserving localization accuracy.
    \item We implement NexusRCL and comprehensively evaluate its effectiveness through extensive experiments on two microservice benchmarks.
\end{enumerate}

\section{Background and Related Work}
\label{sec:background and related work}

The effectiveness of RCL in microservice systems depends on how well a model navigates inherent heterogeneity, which is often discussed ambiguously and modeled incompletely. Heterogeneity arises from both the system structure and diverse monitoring signals, and it is further reflected in the modeling strategies used to integrate and represent these signals. In this section, we decompose heterogeneity in microservice systems into distinct dimensions and review representative RCL methods from the perspective of heterogeneous modeling.

\subsection{Heterogeneity in Microservice Systems}
Microservice systems are inherently heterogeneous, due to architectural decomposition, independent deployment, elastic scaling, and resource management across physical or virtual infrastructures. We decompose system-level heterogeneity into the following dimensions.

\subsubsection{Monitoring Data Heterogeneity}
Microservice systems generate heterogeneous observability signals~\cite{zhang2025failure}, including metrics, logs, and traces. These modalities differ in data structures, semantic abstractions, sampling granularity, temporal continuity, and noise characteristics.


\subsubsection{System Entity Heterogeneity}
Microservice systems inherently exhibit multiple abstraction layers, including hosts and services. Entity heterogeneity exists both within and across these layers. Within the service layer, services differ in functionality, workload, resource usage, invocation frequency, and deployment scale. Within the host layer, hosts vary in hardware configuration, resource capacity, and runtime load. Across layers, hosts and services exhibit distinct monitoring modalities, and even shared modalities (e.g., metrics) differ in semantics and granularity. Fault propagation between layers further amplifies heterogeneity, as discussed in Section~\ref{sec:MOTIVATION}.


\begin{figure}[th]
\setlength{\abovecaptionskip}{0pt}
\setlength{\belowcaptionskip}{0pt}
\centering
\includegraphics[width=0.9\textwidth, trim=-4cm 10.5cm 15cm 0cm, clip]{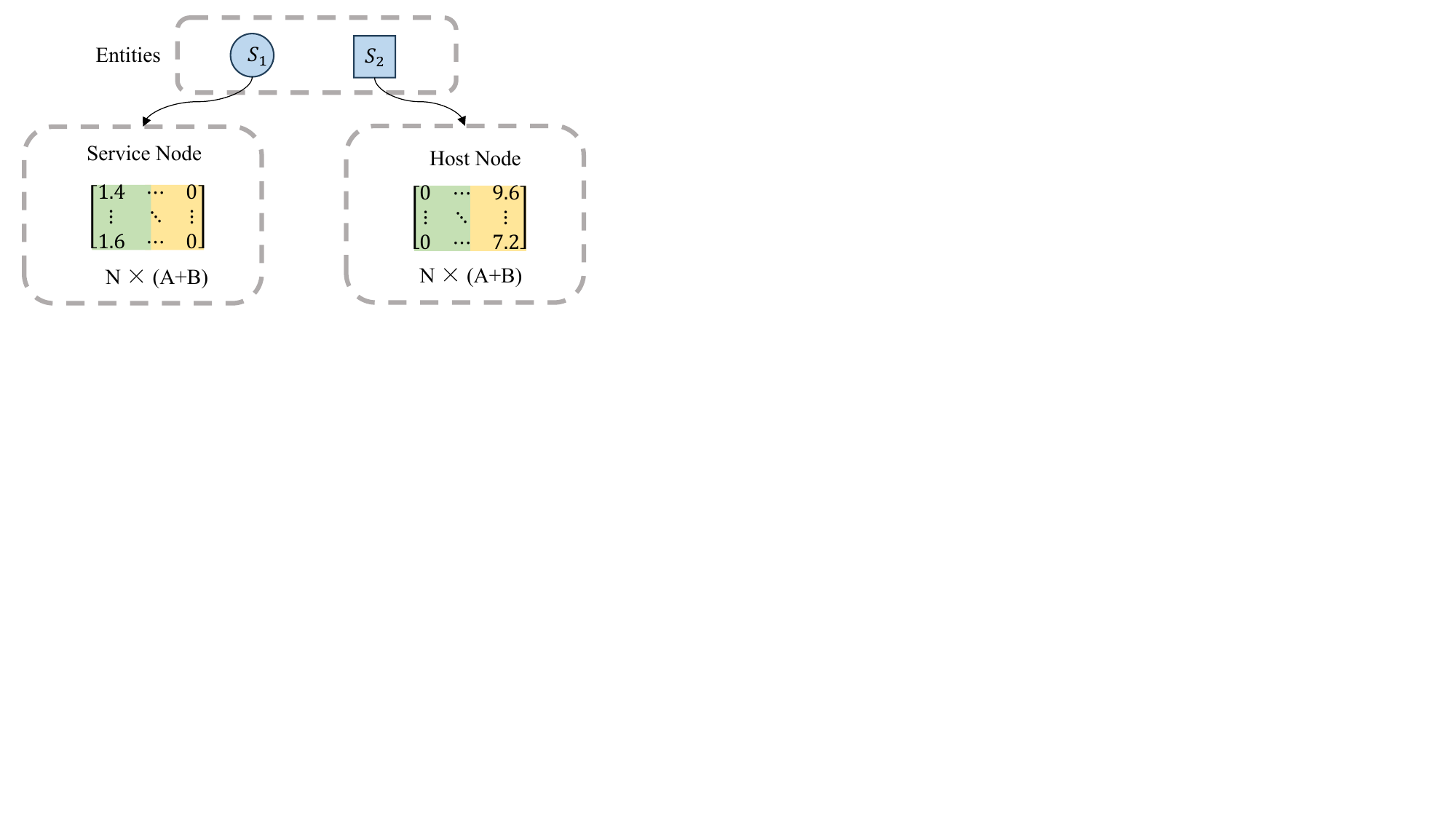}
\caption{Feature Sparsity Caused by Metric-based Fusion.}
\label{fig:motivation2}
\end{figure}


Prior RCL methods address the above heterogeneity mainly through two strategies: (1) data fusion to integrate multiple monitoring modalities, and (2) graph modeling to capture entity-level heterogeneity. In the following, we review representative approaches under these two strategies.

\subsection{Data Fusion}
RCL methods typically integrate metrics, logs, and traces to improve diagnostic coverage and robustness, as fault symptoms may manifest in specific modalities~\cite{hou2021diagnosing}. However, these modalities differ in formats, temporal resolutions, and semantics, making direct fusion difficult~\cite{lee2023eadro,yu2023nezha}. Consequently, existing work primarily adopts two fusion paradigms.

\textbf{Metric-based Fusion.}
This paradigm converts heterogeneous monitoring signals into standardized numerical metrics. Logs are transformed via template matching~\cite{zhang2021cloudrca,wang2024loggt}, while traces contribute aggregated indicators such as latency and success ratio~\cite{tao2024diagnosing}. The missing modalities are filled with placeholders~\cite{sun2024art, sun2025interpretable}, and the resulting metrics are concatenated into a unified feature vector for downstream models.

Despite its simplicity, metric-based fusion has two key limitations:
\begin{itemize}
    \item \textbf{Information dilution due to imbalanced feature volumes.}
    Observability data sources exhibit disparate cardinalities and information densities. For instance, a few high-value metrics may be overshadowed by voluminous, low-information log templates, resulting in information dilution and reduced RCA effectiveness.    
    
    \item \textbf{Feature sparsity.}  
    When data sources are missing, placeholder values create sparse representations (Fig.~\ref{fig:motivation2}), which can hinder model convergence and increase localization latency~\cite{sun2024art,sun2025interpretable}.
\end{itemize}


\textbf{Event-based Data Fusion.}
Instead of forcing heterogeneous signals into numerical metrics, event-based fusion converts metrics, logs, and traces into discrete events that capture anomalous or semantically meaningful behaviors. For example, log events may be generated using Hawkes processes~\cite{lee2023eadro}, while metric anomalies are detected via lightweight detectors~\cite{zhang2021cloudrca}. This approach supports semantic alignment and naturally enables graph-based modeling.


\subsection{Heterogeneous Graph Modeling}



To capture entity-level heterogeneity, recent work introduces heterogeneous graphs with multiple node and edge types. Most methods emphasize service-layer heterogeneity, distinguishing service types or observability artifacts such as calls, traces, or log-derived entities~\cite{tao2024diagnosing,yang2025hg,zhao2024chase}. However, such models often omit explicit host representations and, therefore, cannot localize host-level faults or model cross-layer propagation.

A limited number of works explicitly model both services and hosts~\cite{sun2024art,sun2025interpretable}, but they still rely on unified feature spaces and shared embedding mechanisms across entity types, which can introduce semantic misalignment and inefficient learning due to the fundamentally different observability characteristics of services and hosts.

\begin{table*}[t]
\centering
\caption{Statistics and Entity-layer Fault Propagation Patterns in \textit{HD1} and \textit{HD2}}
\label{tab:dataset_details}
\renewcommand{\arraystretch}{1.2}
\begin{tabular}{|c|cc|cc|c|cc|ccc|}
\hline
\multirow{2}{*}{\textbf{Dataset}} 
& \multicolumn{2}{c|}{\textbf{Entities}} 
& \multicolumn{2}{c|}{\textbf{Failures}} 
& \textbf{Metrics} 
& \multicolumn{2}{c|}{\textbf{Fault Types}} 
& \multicolumn{3}{c|}{\textbf{Propagation}} \\
\cline{2-11}
& \textbf{Tot} & \textbf{Svc/Host} 
& \textbf{Tot} & \textbf{Svc/Host} 
& \textbf{Tot} 
& \textbf{S} & \textbf{H}
& \textbf{NP} & \textbf{Intra} & \textbf{Inter} \\
\hline
\textit{HD1} 
& 46 & 40/6 
& 210 & 168/42 
& 113 
&  10 & 5
& 45 & 23 & 173 \\
\textit{HD2} 
& 38 & 30/8 
& 400 & 318/82 
& 332 
&  15 & 3
& 10 & 81 & 309 \\
\hline
\end{tabular}
\par\vspace{6pt}
\parbox{0.7\textwidth}{
\footnotesize
\textit{Note}: Tot = Total; Svc/Host = Service/Host entities or failures;
NP = No Propagation;
Intra = Propagation among entities of the same type;
Inter = Propagation across different entity types.
}
\end{table*}

\section{Motivation and Challenges}
\label{sec:MOTIVATION}

Consistent with the limitations discussed in Section~\ref{sec:background and related work}, existing RCL methods do not explicitly differentiate distinct forms of heterogeneity. While many approaches claim to be heterogeneity-aware, they typically focus on only a subset of heterogeneity dimensions (e.g., monitoring data heterogeneity or service-centric modeling) and overlook how different types of heterogeneity play inherently unequal roles in fault diagnosis. This one-sided treatment limits the expressive capacity of RCL models and prevents heterogeneous modeling techniques from faithfully capturing real-world fault propagation behaviors.

Among the various forms of heterogeneity present in microservice systems, \emph{system entity heterogeneity} is particularly critical for RCL, yet remains insufficiently explored in prior work. 
In practice, this form of heterogeneity fundamentally shapes how faults propagate across the system, thereby exerting a direct impact on the difficulty and accuracy of RCL. 
We therefore focus on entity-level heterogeneity and examine its implications through the lens of fault propagation.

\subsection{Fault Propagation under Entity-level Heterogeneity}

Entity-level heterogeneity in microservice systems naturally gives rise to heterogeneous fault propagation behavior~\cite{hou2021diagnosing,lee2023eadro,kubernetes2025hpa,tao2024diagnosing}. 
Due to the coexistence of heterogeneous system entities (services and hosts in this paper) and their asymmetric dependencies, faults in real-world microservice systems rarely remain isolated. Instead, they propagate across entities and system layers, often amplifying their impact.


To better understand how entity-level heterogeneity manifests through fault propagation, we performed an empirical analysis on two datasets, HD1 and HD2 (see Section~\ref{sec:Dataset} for details). Faults in these datasets occur at two system layers: service-level faults, which affect individual containers and mainly involve application or container resource anomalies (e.g., k8s container memory load), and host-level faults, which affect the underlying nodes and typically involve hardware or system resource issues (e.g., node memory consumption). Since faults propagate across entities rather than within a single entity, we categorize fault cases based on whether propagation occurs across different entity types.

Specifically, we define three propagation patterns: (i) \emph{No Propagation}, where anomalies remain localized to a single entity; (ii) \emph{Intra-layer Propagation}, where faults spread among entities of the same type (services or hosts); and (iii) \emph{Inter-layer Propagation}, where faults propagate across different entity types (between services and hosts). As summarized in Table~\ref{tab:dataset_details}, the majority of fault cases involve \emph{inter-layer propagation}.


Beyond propagation patterns, we further examine the relationship between fault propagation and fault semantics. We observe that host level fault types exhibit no observable propagation in either dataset. In contrast, among service-level fault types, only I/O-related faults in HD1 (which correspond to underlying file system I/O issues) remain localized, whereas the remaining service fault types consistently trigger fault propagation. These observations suggest that fault propagation is not a stochastic phenomenon, but is tightly coupled with both the semantic nature of the fault and the system entity on which it occurs.

A closer inspection of propagation behavior reveals two recurring principles.
\begin{itemize}
    \item Failures originating from service instances frequently affect their hosting machines, with the propagation likelihood depending on the fault type.
    \item Service failures may further propagate to upstream or downstream services through timeouts, retries, or cascading invocation failures, although such intra-layer propagation is comparatively less frequent.
\end{itemize}

These observations are consistent with operational insights reported in prior studies~\cite{ma2020diagnosing}, which indicate that fault propagation in microservice systems predominantly follows asymmetric cross-layer paths, typically from the service layer to the host layer, while reverse propagation is uncommon.

Taken together, these findings demonstrate that fault propagation behavior is a direct consequence of entity-level heterogeneity in microservice systems. 
Importantly, this form of heterogeneity is fundamentally distinct from the heterogeneity arising from multi-source monitoring data, as it is rooted in asymmetric structural dependencies among system entities rather than in the diversity of observability modalities.

\subsection{Core Motivations}

This analysis leads to our core motivation:
effective RCL in microservice systems requires both embracing system-wide heterogeneity and explicitly prioritizing the forms of heterogeneity that are most diagnostically influential. Failing to make this distinction causes existing RCL approaches to suffer from two fundamental limitations.

\begin{figure*}[!tp] 
    \centering
    \includegraphics[width=\textwidth, trim=0cm 3cm 0cm 3cm, clip]{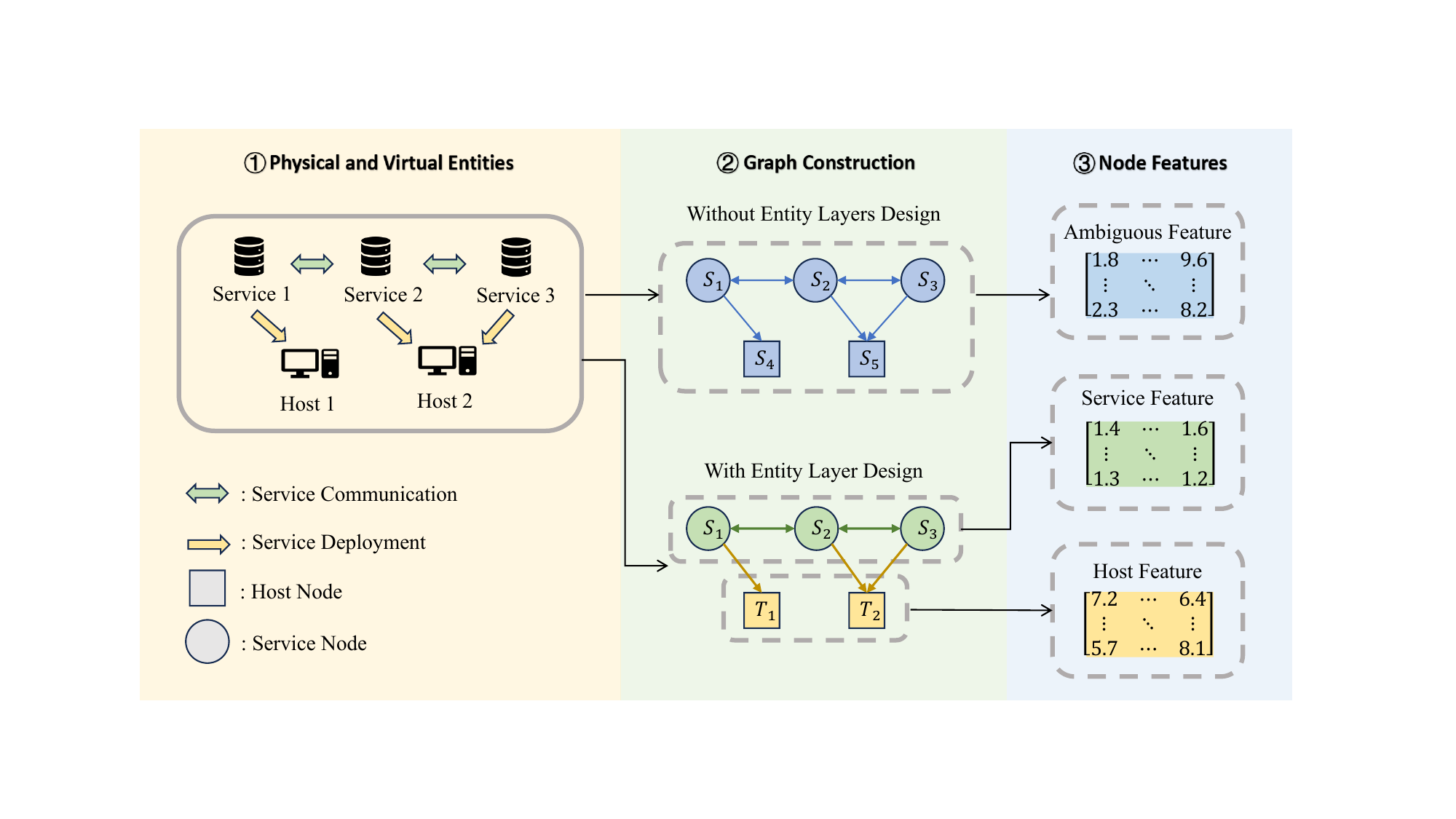}
    \caption{Graph Construction: Without vs. With Explicit Entity Layer Design.}
    \label{fig:motivation1}
\end{figure*}

\textbf{Obscured entity distinctions hinder accurate attribution.}
Many existing methods directly fuse multi-source observability data without explicitly distinguishing between system entities. As a result, service-level and host-level data are entangled in a shared feature space, blurring their semantic boundaries and making it difficult to attribute anomalies to specific components. As illustrated in Fig.~\ref{fig:motivation1}, the absence of an explicit \emph{entity layer} prevents models from capturing the hierarchical structure inherent in microservice architectures.

\textbf{Neglect of asymmetric cross-layer fault propagation.}  
At the structural level, existing RCL methods largely overlook the dominance of cross-layer fault propagation.
Former approaches either assume symmetric interactions among all entities or treat fault propagation uniformly across layers.
However, in practice, failures propagate in a highly asymmetric manner: service-level faults frequently impact hosts, while reverse propagation (host-to-service) is relatively rare.

Together, these limitations motivate the adoption of explicit heterogeneous graph representations, in which services and hosts are modeled as distinct entity types, and typed edges are used to encode invocation, deployment, and resource-dependency relationships.

\subsection{Challenges}
Despite the advantages of heterogeneous graph-based modeling with \emph{explicit layer awareness}, RCL in microservice systems still faces fundamental challenges arising from system dynamics and data scarcity.

\textbf{Challenge 1: Modeling inter-layer heterogeneity under dynamic deployment and complex indicator correlations.}  
Microservice systems are inherently dynamic. While service-to-service interactions are relatively stable, deployment relationships between services and hosts change frequently due to auto-scaling, container migration, and infrastructure churn. Such dynamics violate static topology assumptions and make it difficult to consistently model inter-layer dependencies. Moreover, fault propagation is reflected through heterogeneous observability indicators with complex correlations across entities and layers. Accurately capturing evolving service-host relationships together with these correlated indicators is essential for modeling realistic fault propagation paths in microservice RCL.

\textbf{Challenge 2: Balancing labeling cost and localization performance.}  
High-quality fault labels are scarce and costly to obtain in production microservice systems. Supervised RCL methods typically require large volumes of labeled data to achieve high accuracy, incurring substantial annotation overhead, whereas unsupervised approaches reduce labeling requirements at the cost of strong assumptions that often fail in dynamic real-world environments. Consequently, achieving reliable fault localization performance under limited labeling budgets remains a fundamental challenge.


\subsection{Problem Definition}
\label{sec:problem_definition}

In a microservice system, when a failure occurs within a given time interval, engineers seek to identify the \textit{root cause}, which may be attributed to a service or a host. This task is referred to as Root Cause Localization.

To perform online RCL, the overall time interval is divided into several \textbf{time windows}. A time window is defined as a contiguous interval \([t_s, t_e]\) of fixed length \(e - s\), over which system observability data are aggregated for analysis. Time windows that do not contain any faults are referred to as \textbf{fault-free windows}, whereas those containing at least one fault are referred to as \textbf{failure time windows}, on which RCL is performed.

We model the microservice system as a series of heterogeneous graphs 
\(\{\mathcal{G}_t\}_{t=1}^T\), where each \(\mathcal{G}_t = (\mathcal{V}, \mathcal{E}_t)\) represents the system during a failure time window \(t\). The node set \(\mathcal{V} = \mathcal{V}_{S,t} \cup \mathcal{V}_{H,t}\) includes service nodes \(\mathcal{V}_{S,t}\) and host nodes \(\mathcal{V}_{H,t}\), and \(\mathcal{E}_t\) denotes typed edges. Each graph is associated with node features \(D_t\), capturing event-based observability data.

Within each failure time window \(t\), we formulate RCL as a \textit{ranking problem}, aiming to compute a probability vector 

\begin{equation}
P_t = [p_{1,t}, \dots, p_{M_t,t}] \in [0,1]^{M_t},
\end{equation}

where \(M_t = |\mathcal{V}|\), and each \(p_{i,t}\) represents the likelihood that node \(v_i \in \mathcal{V}\) is the root cause of the failure in time window \(t\).

\section{Methodology}
\label{sec:METHODOLOGY}

\subsection{Overview}
\begin{figure*}[t]
    \centering
    \includegraphics[width=\textwidth, trim=0cm 0cm 0cm 5.5cm, clip]{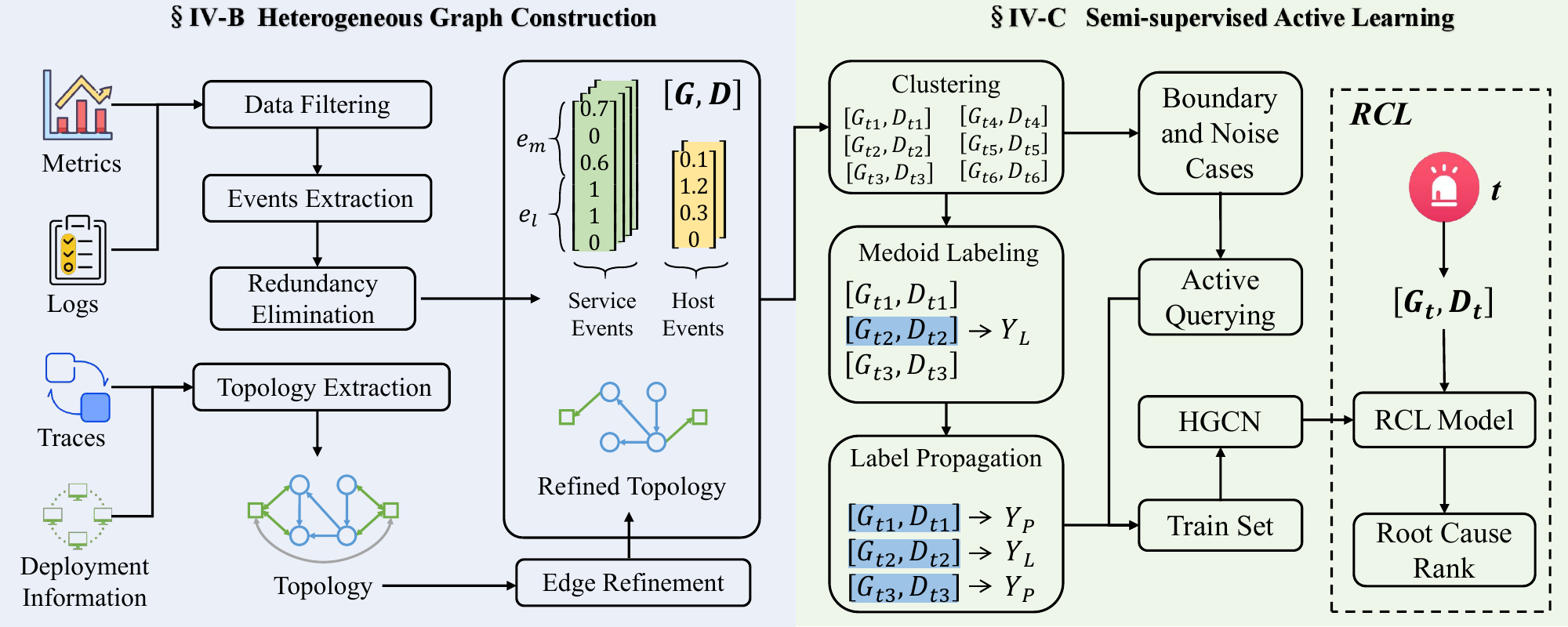}
    \caption{Overview of NexusRCL.}
    \label{fig:NexusRCL_overview}
\end{figure*}

Guided by the empirical findings in Section~\ref{sec:MOTIVATION}, we propose \textbf{NexusRCL}, a framework designed to jointly model data-level and entity-level heterogeneity. As illustrated in Fig.~\ref{fig:NexusRCL_overview}, NexusRCL achieves robust RCL through two tightly integrated components.

\textbf{Heterogeneous Graph Construction}: To capture the asymmetric influences on fault propagation, this module explicitly encodes entity distinctions by formalizing services and hosts as distinct node types within a heterogeneous graph. Rather than prematurely projecting multi-source observability data into a unified feature space, it leverages an \emph{event-based abstraction mechanism} to preserve the intrinsic semantics of diverse signals. This process automatically transforms voluminous raw data into concise and structured representations without human intervention.
    
\textbf{Semi-Supervised Active Learning}: To balance labeling costs with localization performance, this module introduces an iterative supervision paradigm for the generated heterogeneous graphs. It incorporates a three-step active learning strategy, integrating graph embedding clustering, pseudo-label propagation, and uncertainty-based querying. By alternating between expert annotation of critical samples and automated label propagation within semantically coherent clusters, this component produces a high-quality training dataset with minimal manual effort. The resulting model is then deployed for real-time online RCL.

NexusRCL assumes access to the deployment topology information, which is typically recorded in traces and CMDB deployment records. The framework consumes metrics and logs without requiring a fixed schema or format. Each observation only needs to be associated with a specific service or host.

During offline training, NexusRCL incrementally injects supervisory signals via active learning and trains a heterogeneous graph convolutional network (HGCN). The trained model is then deployed online to perform real-time RCL once a fault alarm is triggered.

\subsection{Heterogeneous Graph Construction}
\label{sec:graph_construction}

This module transforms raw multi-source monitoring data into a set of heterogeneous graphs \( (\mathcal{G}_t, D_t)_{t=1}^T \), where each graph \( \mathcal{G}_t = (\mathcal{V}, \mathcal{E}_t) \) captures the system topology during time window \( t \), and \( D_t \) denotes the event features attached to nodes. The construction involves two parallel procedures: (i) \textit{topology extraction}, which determines the graph structure; and (ii) \textit{event extraction}, which derives event features from raw multi-source data.

\subsubsection{Topology Extraction}
\label{sec:topo}

Modern microservice architectures exhibit distinct dynamics at the service and instance levels. 
At the service level, updates (e.g., changes in service interactions or service counts) are deployed cautiously, often using canary releases to ensure system stability~\cite{humble2010continuous,vangala2025blue}. 
In contrast, instance-level dynamics are significantly more volatile due to mechanisms such as Horizontal Pod Autoscaler (HPA), which continuously adjusts the number of instances based on monitoring metrics like CPU utilization or request latency~\cite{kubernetes2025hpa}.

To avoid instance-level volatility, {NexusRCL} abstracts the system topology into a service–host hierarchy, filtering out transient instance churn (see Section~\ref{sec:DISCUSSION}). Each service is represented by a node \(v_s \in V_S\), and edges \(E_{SS}\), \(E_{HH}\) and \(E_{SH}\) capture inter-service invocations, host-level connectivity/co-location, and service deployment relationships, respectively. Specifically, \(E_{SS}\) is derived from distributed traces, \(E_{HH}\) is inferred from physical connectivity and co-location (with redundant bidirectional edges removed when a directed \(E_{SS}\) edge exists), and \(E_{SH}\) is obtained from deployment metadata and oriented from service to host to reflect fault propagation.

\subsubsection{Event Extraction}

Given the graph topology, we encode event-specific features as node attributes. This event-based abstraction mechanism enables joint modeling of structural and contextual information.

\textbf{Data Filtering.}  
We first denoise the raw data to remove irrelevant or unstable signals. For metric selection, we discard time series with either low variance (e.g., those exhibiting no fluctuations at all) or excessive fluctuations (mean variation exceeding twice the global average). The mean variation over a sliding window \( [t-W, t] \) is defined as:

\begin{equation}
\text{Mean Variation} = \frac{1}{W} \sum_{t_i \in [t-W, t]} |m(t_i) - m(t_{i-1})|.
\end{equation}

For logs, we apply regular expressions to remove low-information tokens such as numeric IDs and IP addresses, retaining meaningful semantic content for subsequent processing. For logs with severity levels, we filter out \texttt{INFO}-level logs and retain only \texttt{WARNING} and \texttt{ERROR} logs to focus on critical events.

\textbf{Metric Event Extraction.}  
We apply the \( n \)-sigma rule~\cite{zhang2021cloudrca} to detect metric deviations. Given a metric series \( m(t) \), the event signal is computed as:

\begin{equation}
e_m(t) = m(t) - \mu_{[t-W, t]} - n \cdot \sigma_{[t-W, t]},
\end{equation}

where \( \mu \) and \( \sigma \) are the local mean and standard deviation over window \( [t-W, t] \), and \( n \) is a sensitivity parameter. This formulation preserves both the magnitude and direction of anomalies, enabling detection beyond binary flags.

Existing localization-oriented anomaly detectors often trade accuracy for efficiency, while more sophisticated methods~\cite{ma2020automap,liu2021microhecl} incur higher overhead. We adopt the lightweight \( n \)-sigma method to balance detection quality and computational cost.

\textbf{Log Event Extraction.}  
Raw logs are transformed into semantic vector embeddings using pre-trained models such as SentenceTransformer~\cite{reimers2019sentence}, followed by clustering via DBSCAN~\cite{ester1996density} to identify recurring log event patterns. Let \( K \) denote the number of resulting clusters. For each time window \( t \), we construct a binary vector \( e_l(t) \in \{0, 1\}^K \), where the \( k \)-th element is set to 1 if at least one log within the window is assigned to cluster \( k \):

\begin{equation}
e_l^{(k)}(t) =
\begin{cases}
1, & \text{if any log in } t \text{ belongs to cluster } k, \\
0, & \text{otherwise}.
\end{cases}
\label{eq:el_def}
\end{equation}

Logs identified as noise by DBSCAN are excluded. This multi-hot encoding effectively captures the presence of diverse log event types within each time window and constitutes the log-based component of the node data.

\textbf{Topology Change Event.}
In addition to metric and log events, we incorporate \textbf{topology change events} to capture dynamic variations in service-to-host associations, which are critical for RCL when topology shifts occur due to faults.

Topology change events are derived from service-to-host edge matrices at consecutive time steps. The service-to-host edge matrix is \(E_{SH}(t) \in \{0,1\}^{|S| \times |H|}\) for time step \(t\). We compute the difference between the matrices at \(t\) and \(t-1\):

\begin{equation}
\Delta E_{SH}(t) = E_{SH}(t) - E_{SH}(t-1)
\end{equation}

For each service node \(v_s^i\), the topology change feature \(e_{c}^i(t)\) is obtained by aggregating the absolute differences over all associated hosts:

\begin{equation}
e_{c}^i(t) = \sum_{j=1}^{|H|} \left| \Delta E_{SH}(i,j)(t) \right|
\end{equation}

This event signal provides contextual information for RCL under dynamic service-to-host mappings, enabling the model to account for topology shifts.

\textbf{Event Fusion.}  
We fuse above signals to generate the final node-level event vectors. For \emph{service nodes} with all three event signals:

\begin{equation}
x_s(t) = [e_m(t) \, \| \, e_l(t) \|\, e_c(t)] \in \mathbb{R}^{A + K + 1},
\end{equation}

where \( A \) is the number of service-specific metric dimensions, and \( K \) is the number of log clusters.

For \emph{host nodes}, which often lack meaningful log data:

\begin{equation}
x_h(t) = e_m(t) \in \mathbb{R}^{B},
\end{equation}

where \( B \) is the number of host-specific metric dimensions, typically \( B \neq A \). This design ensures that each node type is represented by features aligned with its monitoring characteristics.

\subsubsection{Finetuning Design}
With the previously obtained events and topology, we proceed with the following two steps to derive the final graphs.

\textbf{Redundancy Elimination.}  
To retain only informative signals in the node data \( D_t \), we eliminate redundant events. For metrics, periodic patterns (e.g., regular CPU usage peaks) are removed using Seasonal-Trend decomposition based on Loess (STL)~\cite{rb1990stl}. The period required by STL is preset, and in our experiments it is set to one hour. A periodic baseline \( p(t) \) is estimated and subtracted from the raw series to yield residuals \( m'(t) = m(t) - p(t) \), upon which anomalies are re-evaluated. For logs, we discard event clusters with distributed timestamps, identified via low temporal variance \( \text{Var}(f_e(t)) < \delta \), as they are unlikely to correlate with failures.

After event extraction, all metric and log events within each time window \( t \) are concatenated into a unified event vector per node, forming the node feature set \( D_t \). 

\textbf{Edge Refinement.}  
To enhance the structural quality of the heterogeneous graph \( \mathcal{G} \), we refine its edge set \( \mathcal{E} \) through correlation-based pruning and directionality enforcement. For each candidate edge, we compute the Pearson correlation between the event frequencies \( f_i(t) \) and \( f_j(t) \) of the connected nodes. Edges with correlation below a threshold \( \gamma \) are removed, eliminating weak or irrelevant relationships:

\begin{equation}
\text{Corr}(f_i(t), f_j(t)) = \frac{\text{Cov}(f_i(t), f_j(t))}{\sigma_{f_i} \cdot \sigma_{f_j}} < \gamma.
\end{equation}

In addition, we apply directionality constraints inspired by operational domain knowledge~\cite{ma2020diagnosing}, which indicate that faults predominantly propagate across layers, from service nodes to host nodes. Consequently, all bidirectional edges between the service and host nodes (\( E_{SH} \)) are converted to unidirectional edges that point from the service to the host, aligning the graph topology with empirical failure propagation patterns.

\textbf{Final Heterogeneous Graph.}  
Following these refinement steps, {NexusRCL} constructs a set of heterogeneous graphs \( \{(\mathcal{G}_t, D_t)\}_{t=1}^T \), where each tuple corresponds to a time window \( t \), representing an individual fault case. Specifically:

\begin{itemize}
    \item \( \mathcal{G}_t = (\mathcal{V}, \mathcal{E}_t) \), where \( \mathcal{V} = V_S \cup V_H \) denotes the set of service and host nodes, and \( \mathcal{E}_t = E_{SS} \cup E_{SH} \cup E_{HH} \) represents service-to-service, service-to-host, and host-to-host relationships, respectively;
    \item \( D_t = \{X^S_t, X^H_t\} \), where \( X^S_t = \{X_{i,t}^S \mid v_i \in V_S\} \) comprises feature vectors for service nodes (derived from metrics, logs, and traces), and \( X^H_t = \{X_{j,t}^H \mid v_j \in V_H\} \) includes feature vectors for host nodes (derived from metrics), with differing vector lengths due to distinct node characteristics.
\end{itemize}

At this stage, heterogeneous graphs \( (\mathcal{G}_t, D_t)_{t=1}^T \) are constructed to capture the topological structure and event-driven node characteristics. However, they remain unlabeled until the active learning phase (Section~\ref{sec:active_learning}), where label information is introduced.

\subsection{Semi-Supervised Active Learning}
\label{sec:active_learning}

Although heterogeneous graphs provide strong expressive power, obtaining labeled fault cases in microservice systems remains costly and labor-intensive. 
{NexusRCL} addresses this challenge by coupling heterogeneous graph modeling with a semi-supervised active learning strategy that balances label efficiency and model robustness.

Each representation \( (\mathcal{G}_t, D_t) \), corresponding to a fault case in time window \( t \), is initially unlabeled. We define a one-hot label vector \( Y_t = \{y_{i,t}\}_{i=1}^{|\mathcal{V}|} \) per representation, where $y_{i,t} \in \{0, 1\}$ indicates whether node $v_i \in \mathcal{V}$ is a root cause. Labels are derived from two sources to form a partially labeled dataset:

\begin{itemize}
    \item \textbf{Human-annotated Labels} (\( Y_t^L \)): Cluster medoids (Step 2) and actively queried nodes, including boundary nodes and noise points (Step 3), are manually annotated by SREs to produce trusted labels.
    \item \textbf{Pseudo-Labels} (\( Y_t^P \)): All clustered nodes inherit their cluster medoid's label via label propagation, yielding pseudo-labels.
\end{itemize}

The label set \( Y_t = Y_t^L \cup Y_t^P \) is associated with \( (\mathcal{G}_t, D_t) \), forming a partially labeled dataset \( \{(\mathcal{G}_t, D_t, Y_t^L, Y_t^P)\}_{t=1}^T \) for RCL training. Active learning iteratively queries high-uncertainty nodes to enhance label efficiency, with the pipeline comprising: (1) graph-aware embedding and clustering, (2) medoid labeling and propagation, and (3) boundary and noise case refinement.

\textbf{Step 1: Graph-Aware Embedding and Clustering}

We employ a HGCN to compute embeddings for each representation \( (\mathcal{G}_t, D_t) \):

\begin{equation}
Z_t = \text{HGCN}(\mathcal{G}_t, D_t),
\end{equation}

where \( Z_t = \{Z_{i,t} \mid v_i \in \mathcal{V}\} \) denotes the unified embedding space. The HGCN is trained with a self-supervised reconstruction objective to capture both structural dependencies and attribute semantics.

To partition these embeddings, we adopt DBSCAN, which detects clusters of arbitrary shapes, separates dense regions from sparse noise, and does not require specifying the number of clusters. As shown in Step 2, DBSCAN outperforms alternatives such as K-means and hierarchical clustering for our setting. The resulting clusters \( \{C_k\} \) and noise points form the basis for downstream labeling.

This design also removes the need to assume that every time window contains a fault. Fault-free windows, which are abundant in practice, are consistently grouped into a dominant cluster that serves as a baseline class and helps filter out false alarms. Since normal cases are easily separable, they are excluded from active learning, reducing labeling overhead. Only when fault-free data are unavailable would one need to assume that every window corresponds to a genuine fault.

\textbf{Step 2: Medoid Labeling and Label Propagation}

For each cluster \( C_k \), we select a medoid \( m_k \) for SRE labeling to generate trusted labels \( Y_t^L \) and propagate them to form pseudo-labels \( Y_t^P \). The medoid minimizes the total Euclidean distance to other nodes in \( C_k \) within \( Z_t \):

\begin{equation}
m_k = \arg\min_{v_i \in C_k} \sum_{v_j \in C_k} \| Z_{i,t} - Z_{j,t} \|_2.
\end{equation}

SREs annotate the original representation \( (\mathcal{G}_t, D_t) \) corresponding to the medoid, producing a trusted label. This label is propagated to the samples corresponding to all other embeddings in \( C_k \):

\[
Y_{s}^P = Y_{m_k}^L, \quad \forall Z_{s} \in C_k,
\]

where \( Y_s^P \in \{0, 1\}^{|\mathcal{V}_s|} \) is the pseudo-label vector for each representation. This label propagation efficiently generates pseudo-labels for semi-supervised RCL training.  

\begin{table}[h]
\centering
\caption{Clustering Algorithm Accuracy on Datasets \textit{HD1} and \textit{HD2}}
\label{tab:clustering_accuracy}
\renewcommand{\arraystretch}{1.2}
\setlength{\tabcolsep}{12pt}
\begin{tabular}{|c|c|c|}
\hline
\textbf{Method} & \textbf{HD1 Accuracy} & \textbf{HD2 Accuracy} \\
\hline
DBSCAN          & 0.86                  & 0.64                  \\
K-means         & 0.71                  & 0.56                  \\
Hierarchical    & 0.68                  & 0.53                  \\
\hline
\end{tabular}
\end{table}

To evaluate clustering effectiveness, we assess pseudo-label quality. For each cluster, 15 non-medoid nodes are randomly sampled and manually labeled by SREs to obtain ground-truth labels. Accuracy is the proportion of pseudo-labels \( Y_t^P \) matching these ground-truth labels, averaged over five runs (Table~\ref{tab:clustering_accuracy}). DBSCAN achieves the highest accuracy (0.86 on \textit{HD1}, 0.64 on \textit{HD2}), outperforming K-means~\cite{macqueen1967some} and Hierarchical Clustering~\cite{ward1963hierarchical}, confirming its suitability for our task.  

\textbf{Step 3: Boundary and Noise Case Refinement}

To address nodes deviating from their cluster's dominant semantics, we actively query boundary nodes and DBSCAN noise points for manual labeling. For each cluster \( C_k \), we select the top-\( K \) nodes with the largest Euclidean distance to the medoid \( m_k \):

\begin{equation}
\mathcal{B}_k = \text{TopK}_{v_i \in C_k} \left( \| Z_{i,t} - Z_{m_k,t} \|_2 \right).
\end{equation}

Noise points identified by DBSCAN are also included as query cases, as they may represent rare root causes or outliers. By prioritizing these high-uncertainty cases, we direct manual labeling efforts toward samples that are most likely to influence the model's decision boundaries. These selected nodes are submitted to SREs for annotation, resulting in high-confidence labels that are incorporated into the training set. This refinement stage enhances the model's discriminative capacity and supports better generalization in the semi-supervised setting.


In practice, NexusRCL adopts a sequential allocation scheme for labeling. 
First, all cluster medoids are labeled, as their number is fixed and they provide a representative overview of each cluster. 
Next, any remaining labeling budget is used to annotate high-uncertainty samples in a prioritized manner: DBSCAN-identified noise points are labeled first, followed by boundary samples within clusters. 
This sequential strategy ensures that the most informative samples are labeled in order of priority without requiring a predetermined proportion between cluster medoids, noise, and boundary points.


\subsection{RCL Model Training}

Using the partially labeled dataset, NexusRCL trains an HGCN-based scoring function

\begin{equation}
F : (\mathcal{G}_t, D_t) \rightarrow [0,1]^{|\mathcal{V}|},
\end{equation}

which assigns each node a likelihood of being the root cause.
The model is optimized using cross-entropy loss over human-verified labels.
Pseudo-labels are incorporated only when ground-truth annotations are unavailable.

We note that NexusRCL does \emph{not} assume pseudo-labels to be accurate.
Incorrect pseudo-labels are unavoidable unless all fault instances are manually annotated.
To ensure label reliability, NexusRCL prioritizes human-verified annotations selected via uncertainty-based active learning, rather than relying solely on label propagation.
As demonstrated in Section~\ref{sec:EXPERIMENT}, actively selected labels consistently outperform randomly chosen annotations.

At the same time, incorporating pseudo-labels yields better performance than training on a limited set of human-labeled samples alone by providing broader behavioral coverage. We acknowledge that pseudo-labels do not guarantee performance improvement in all cases. This limitation is inherent to semi-supervised learning and is discussed further in Section~\ref{sec:DISCUSSION}.

By jointly integrating entity-aware graph modeling, uncertainty-driven active learning, and cautious use of pseudo-labels, NexusRCL adheres to its central design principle:
\emph{prioritizing diagnostically meaningful heterogeneity is more effective than indiscriminately modeling all forms of heterogeneity}.

\section{Experiment}
\label{sec:EXPERIMENT}

\begin{table*}[t]
\centering
\caption{RCL Performance on \textit{HD1} and \textit{HD2}}
\label{tab:rcl_performance}
\renewcommand{\arraystretch}{1.2}
\begin{tabular}{|c|cccc|cccc|}
\hline
\multirow{2}{*}{\textbf{Method}} & \multicolumn{4}{c|}{\textbf{\textit{HD1}}} & \multicolumn{4}{c|}{\textbf{\textit{HD2}}} \\
\cline{2-9}
 & \textbf{A@1} & \textbf{A@3} & \textbf{A@5} & \textbf{Avg A@5} & \textbf{A@1} & \textbf{A@3} & \textbf{A@5} & \textbf{Avg A@5} \\
\hline
NexusRCL & \textbf{82.50} & \textbf{90.20} & \textbf{92.10} & \textbf{88.15} & \textbf{68.75} & \textbf{74.30} & \textbf{78.50} & \textbf{73.50} \\
\textit{CausalRCA}~\cite{xin2023causalrca} & 17.80 & 38.50 & 53.20 & 36.80 & 18.90 & 42.60 & 60.40 & 40.80 \\
\textit{ART}~\cite{sun2024art} & 51.20 & 54.80 & 57.60 & 54.50 & 13.20 & 25.00 & 48.30 & 28.90 \\
\textit{Eadro}~\cite{lee2023eadro} & 20.30 & 38.90 & 54.70 & 38.10 & 7.80 & 17.90 & 36.60 & 20.80 \\
\textit{DiagFusion}~\cite{zhang2023robust} & 52.10 & 82.70 & 89.60 & 74.80 & 14.90 & 36.00 & 46.30 & 32.50 \\
\textit{DejaVu}~\cite{li2022actionable} & 31.50 & 43.20 & 58.40 & 44.50 & 4.50 & 12.80 & 16.70 & 11.60 \\
\hline
\end{tabular}
\end{table*}

We evaluate NexusRCL to address three research questions:

\textbf{RQ1}: How effective is NexusRCL in RCL within microservice architectures in terms of accuracy and efficiency?

\textbf{RQ2}: To what extent do NexusRCL's individual components contribute to its overall performance in RCL?

\textbf{RQ3}: How do NexusRCL's key hyperparameters affect its performance?

\subsection{Experimental Setup}

\subsubsection{Datasets}
\label{sec:Dataset}

We evaluate NexusRCL on two benchmark heterogeneous microservice datasets, HD1 and HD2. Summary statistics are provided in Table~\ref{tab:dataset_details}. Both datasets will be publicly released through the repositories accompanying our source code\footnote{\label{note:nexusrcl}\url{https://github.com/molujia/NexusRCL}}.  

The HD1 dataset is derived from a simulated e-commerce microservice system deployed on the CCB cloud. It is adapted from the Hipster Shop project~\cite{hipster_shop}, consisting of 10 services deployed across 6 virtual machines, with 4 instances per service. Fault scenarios are designed to reflect realistic failure conditions. The dataset provides dynamic service topology, traces, multi-layer metrics (container, OS, JVM), and logs. The raw data include 56 service-level and 57 host-level metrics, which are reduced to 20 and 55 after filtering.

The HD2 dataset is based on the Online Boutique project~\cite{microservices_demo}, running on 8 virtual machines with 10 services and 3 instances per service (30 instances in total). Faults are injected at both service and node levels, with complete topology, trace, metric, and log data. The raw dataset contains 45 service-level and 287 host-level metrics, and 40 service-level and 167 host-level metrics are retained after filtering.

Both datasets cover common fault types in microservice systems~\cite{zhang2025failure}, including container-level failures (e.g., hardware and network faults) and node-level failures (e.g., CPU, disk, and memory faults). Although the number of fault types differs across layers, the occurrences of fault types within each layer are relatively balanced.

For both datasets, data are split chronologically, using the earliest 70\% for training and the remaining 30\% for testing. Ground-truth labels are obtained from injection logs, while NexusRCL uses only a limited portion of them during training under its active learning strategy.

\subsubsection{Baseline Approaches}
We compare NexusRCL with five state-of-the-art RCL methods for microservice architectures, organized by their underlying strategies.

\begin{itemize}
    \item \textit{DiagFusion}~\cite{zhang2023robust}: a diagnostic framework that integrates multi-source observability data using machine learning techniques.    \item \textit{CausalRCA}~\cite{xin2023causalrca}: a recent causal inference (CI)-based method that builds upon earlier CI approaches such as RCD~\cite{ikram2022root} and CIRCA~\cite{li2022causal}.    \item \textit{ART}~\cite{sun2024art}: a heterogeneous graph-based approach that employs multi-task learning for anomaly detection and localization.    \item \textit{Eadro}~\cite{lee2023eadro}: a supervised method that leverages Graph Attention Networks (GATs) to model inter-component dependencies for RCL.    \item \textit{DejaVu}~\cite{li2022actionable}: a supervised machine learning method that provides actionable fault localization.  
\end{itemize}  


\subsubsection{Evaluation Metrics}
We evaluate RCL performance using the Top-\( K \) metric, which measures the probability that the true root cause appears within the top-\( K \) predicted candidates:

\begin{equation}
\text{Top-}K = \frac{1}{N} \sum_{i=1}^{N} \mathbb{I}(g_i \in P_{i,[1:K]}),
\end{equation}

where \( g_i \) is the ground truth root cause for the \( i \)-th failure, \( P_{i,[1:K]} \) denotes the top-\( K \) predicted candidates, \( N \) is the total number of failures, and \( \mathbb{I}(\cdot) \) is an indicator function (1 if true, 0 otherwise). We report A@1, A@3, and A@5 to evaluate localization accuracy across ranks, and compute the Average A@5 as \( \text{Average A@5} = \frac{1}{5} \sum_{K=1}^{5} \text{A@K} \) for a comprehensive performance measure.

\subsubsection{Implementation}
NexusRCL is implemented using Python 3.8, PyTorch 2.4.0 with CUDA 12.8, and DGL 0.9.1. Experiments are conducted on a server with 16 CPU cores, 376GB RAM, and NVIDIA RTX A6000 GPUs. HGCN training uses GPUs, while CPUs handle preprocessing. Results are averaged over five runs to ensure robustness.


\begin{table*}[t]
\centering
\caption{Ablation Study of NexusRCL's Components on \textit{HD1} and \textit{HD2}
}
\label{tab:ablation_study}
\renewcommand{\arraystretch}{1.2}
\begin{tabular}{|c|cccc|cccc|}
\hline
\multirow{2}{*}{\textbf{Variant}} & \multicolumn{4}{c|}{\textbf{\textit{HD1}}} & \multicolumn{4}{c|}{\textbf{\textit{HD2}}} \\
\cline{2-9}
 & \textbf{A@1} & \textbf{A@3} & \textbf{A@5} & \textbf{Avg A@5} & \textbf{A@1} & \textbf{A@3} & \textbf{A@5} & \textbf{Avg A@5} \\
\hline
NexusRCL & \textbf{82.50} & \textbf{90.20} & \textbf{92.10} & \textbf{88.15} & \textbf{68.75} & \textbf{74.30} & \textbf{78.50} & \textbf{73.50} \\
A & 75.80 & 81.50 & 88.40 & 84.60 & 19.80 & 44.50 & 63.80 & 43.30\\
B & 80.90 & 86.70 & 89.30 & 85.40 & 62.30 & 67.90 & 72.10 & 67.10\\
C & 55.60 & 79.80 & 84.50 & 73.20 & 23.40 & 38.10 & 53.90 & 38.50 \\
\hline
\end{tabular}
\end{table*}

\begin{table}[htbp]
\centering
\caption{Training (Offline) and RCL (Online) Times for All Methods (in Seconds)}
\label{tab:time_comparison}
\renewcommand{\arraystretch}{1.2}
\begin{tabular}{|c|cc|cc|}
\hline
\multirow{2}{*}{\textbf{Method}} & \multicolumn{2}{c|}{\textbf{\textit{HD1}}} & \multicolumn{2}{c|}{\textbf{\textit{HD2}}} \\
\cline{2-5}
 & \textbf{Offline} & \textbf{Online} & \textbf{Offline} & \textbf{Online} \\
\hline
NexusRCL & 71.0 & 2.50 & 121.4 & 1.20 \\
\textit{CausalRCA}~\cite{xin2023causalrca} & 735.2 & \textbf{0.62} & 542.8 & \textbf{0.26} \\
\textit{ART}~\cite{sun2024art} & \textbf{57.4} & 1.10 & \textbf{77.2} & 1.30 \\
\textit{Eadro}~\cite{lee2023eadro} & 1235.9 & 1.80 & 2240.6 & 3.30 \\
\textit{DiagFusion}~\cite{zhang2023robust} & 1730.5 & 4.60 & 1751.4 & 3.90 \\
\textit{DejaVu}~\cite{li2022actionable} & 148.4 & 5.90 & 135.0 & 5.50 \\
\hline
\end{tabular}
\end{table}

\subsection{RQ1: Overall Performance}
As shown in Table~\ref{tab:rcl_performance}, NexusRCL and most baseline methods exhibit lower performance on \textit{HD2} compared to \textit{HD1}. This discrepancy can be attributed to two main factors: (A) \textit{HD2} is significantly larger and contains a greater number of fault cases, which presents challenges to methods in handling large-scale data; (B) \textit{HD2} includes a richer set of metrics, increasing the difficulty of effectively extracting and utilizing informative signals.

Among the baselines, \textit{ART} and \textit{DiagFusion} perform relatively well on \textit{HD1}, but they rely on metric-based fusion that aggregates service- and host-derived metrics into a single feature vector. This approach biases model attention toward modalities with more metrics, regardless of fault relevance. For example, on \textit{HD2}, each host contributes 287 metrics while each service provides only 45, yet host faults (72) are far less frequent than service faults (408). Consequently, ART and DiagFusion tend to overemphasize hosts, yielding higher accuracy for host faults but poorer performance on service faults, especially under imbalanced metric or fault distributions.

\textit{DejaVu} shares the same limitation of ignoring microservice heterogeneity and performs poorly on both datasets. \textit{CausalRCA} improves on \textit{HD2} compared to \textit{HD1} due to its CI-based framework being less sensitive to data scale. \textit{Eadro} performs poorly on both datasets, as it cannot effectively handle faults occurring at both host and service levels.

NexusRCL achieves the best accuracy across both datasets. On \textit{HD1}, it reaches A@1 of 82.50\%, outperforming \textit{ART} (51.20\%) and \textit{DiagFusion} (52.10\%) by over 30 points, with an inference time of 2.50 s. On \textit{HD2}, it achieves A@1 of 68.75\%, surpassing \textit{CausalRCA} (18.90\%) by nearly 50 points, with an inference time of 1.20 s. It also attains the highest Average A@5 on both datasets (88.15\% on \textit{HD1}, 73.50\% on \textit{HD2}), demonstrating robustness in large-scale heterogeneous settings.

Table~\ref{tab:time_comparison} shows that NexusRCL offers a favorable accuracy–efficiency trade-off. Offline training takes 71.0 s on \textit{HD1} and 121.4 s on \textit{HD2}, faster than \textit{CausalRCA} and \textit{DiagFusion} but slower than \textit{ART}. Online inference remains practical, completing within 2.50 s on \textit{HD1} and 1.20 s on \textit{HD2}.

Overall, NexusRCL provides the best balance between accuracy and efficiency, maintaining high performance with reasonable computational cost across datasets of different scales and heterogeneity.

\subsection{RQ2: Ablation Study}

We evaluate the contributions of NexusRCL's core components using three ablated variants: \textbf{A} (no Edge Refinement), \textbf{B} (no Active Learning, replaced by random labeling), and \textbf{C} (Homogeneous Graphs, replacing heterogeneous graphs and omitting layer concepts).

Table~\ref{tab:ablation_study} shows that all components improve RCL performance, with varying importance across datasets. On \textit{HD1}, removing any component reduces A@1, with edge refinement and heterogeneous graphs having larger impacts (82.50\% to \$ 75.80\% and 55.60\%, respectively). On \textit{HD2}, the heterogeneous graph representation is most critical, followed by edge refinement, while active learning yields a smaller but non-negligible gain.

These results confirm that each component contributes distinct benefits and that their combination yields a balanced architecture adaptable to diverse microservice environments.

\begin{figure*}[ht]
    \centering
    \includegraphics[width=0.8\textwidth]{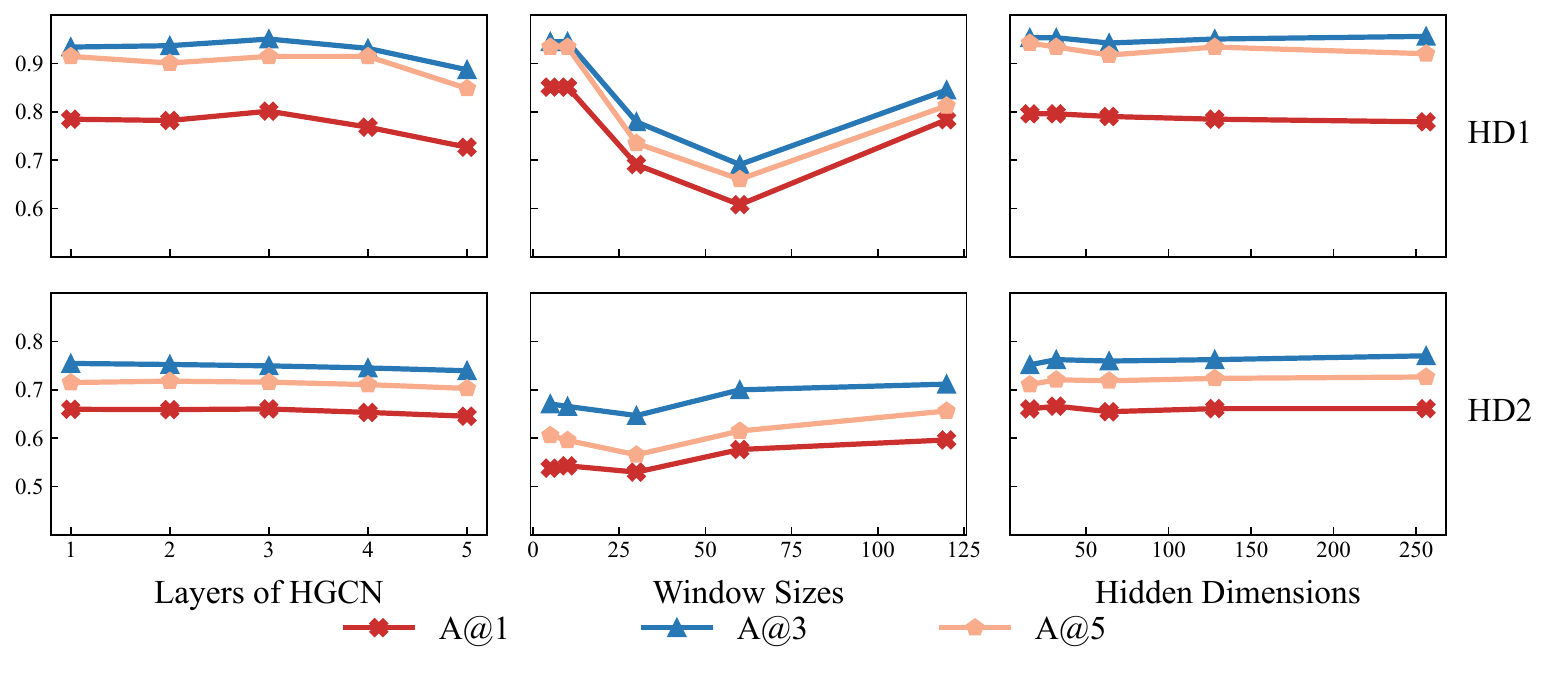}
    \caption{Hyperparameter Sensitivity of NexusRCL on \textit{HD1} and \textit{HD2}.}
    \label{fig:hyperparam_sensitivity}
\end{figure*}

\subsection{RQ3: Hyperparameter Sensitivity}

We evaluate the sensitivity of NexusRCL to three hyperparameters: the number of HGCN layers, the hidden dimension size, and the event aggregation window. As shown in Fig.~\ref{fig:hyperparam_sensitivity}, NexusRCL is relatively robust to changes in layer count and hidden dimension on both \textit{HD1} and \textit{HD2}.

The window size has dataset-dependent effects. In \textit{HD2}, larger windows generally improve performance, while in \textit{HD1} the curve is elbow-shaped, with medium-sized windows performing worst. This may be related to hourly periodic fluctuations in \textit{HD1}, where small or large windows either ignore or smooth out fluctuations, whereas medium windows may misclassify them. We leave a deeper investigation of this phenomenon to future work.

Overall, the best performance requires \textbf{grid search} over the joint hyperparameter space, as optimal values are not independent.

\section{Discussion}
\label{sec:DISCUSSION}

Despite encouraging results, several threats to validity should be acknowledged.

\textbf{Service-level dynamics.}
NexusRCL is robust to instance- and host-level fluctuations and service removals, but it cannot directly handle the emergence of entirely new services. New services are unseen during training, and their introduction may change interaction patterns in ways the model cannot anticipate. Therefore, when the service set changes significantly, retraining is required to adapt to the updated topology.

\textbf{Pseudo-label usage.}
NexusRCL does not assume pseudo-labels to be error-free. To mitigate label noise, we use uncertainty-based active learning to prioritize low-confidence samples for human verification and progressively replace unreliable pseudo-labels. All labels used in evaluation are human-verified. Pseudo-labeling is most beneficial in large-scale settings where full annotation is infeasible; for small datasets, manual labeling may be preferable due to the risk of noise.

\section{Conclusion}
In this paper, we systematically investigate the impact of entity-level heterogeneity on RCL within microservice systems. Our empirical findings reveal that fault propagation is highly asymmetric and dominated by cross-layer interactions between services and hosts. 

To address these characteristics, we propose {NexusRCL}, a semi-supervised framework that explicitly models services and hosts as distinct node types within a heterogeneous graph. By integrating an event-based abstraction mechanism with a three-step active learning strategy, NexusRCL effectively captures complex diagnostic patterns while significantly reducing manual labeling costs. Extensive evaluations on industrial datasets demonstrate that NexusRCL achieves a substantial performance gain, improving Top-1 accuracy by up to 49.85\% over state-of-the-art methods.

\bibliographystyle{unsrt}  
\bibliography{references}

\end{document}